\documentstyle[12pt]{article}

\newcommand{\bra}{\langle}
\newcommand{\ket}{\rangle}
\newcommand{\cH}{{\cal H}}

\begin{document}

\title{\large \bf Quantum degrees of freedom of a region of spacetime\footnote{Proceedings of the
 XXVIII Spanish Relativity Meeting,  Oviedo Spain
September 6-10, 2005}}

\author{\normalsize Federico Piazza \\
\normalsize \it Institute of Cosmology and Gravitation, University of Portsmouth, UK}

\date{}

\maketitle 

\vspace{-7cm}


\vspace{6cm}
\begin{abstract}
The holographic principle and the thermodynamics of de Sitter space suggest that the total number of fundamental degrees of freedom associated with any
finite-volume region of space may be finite. The naive picture of a short distance
cut-off, however, is hardly compatible with the dynamical properties of spacetime, let alone with Lorentz invariance. Considering the regions of space just as general ``subsystems'' may help clarifying this problem. In usual QFT the regions of space are, in fact, associated with a tensor product
decomposition of the total Hilbert space into ``subsystems'', but such a decomposition is given a priori and the fundamental degrees of freedom are labelled, already from the beginning, by the spacetime points. We suggest a new strategy to identify ``localized regions'' as ``subsystems'' in a way which is intrinsic to the total Hilbert-space dynamics of the quantum state of the fields.

\end{abstract}


%
\setcounter{footnote}{0} \setcounter{page}{1}
\setcounter{section}{0} \setcounter{subsection}{0}
\setcounter{subsubsection}{0}

\newpage
In quantum field theory (QFT), independent degrees of freedom are associated with each localized and space-like separated region of spacetime. In the presence of a UV cut-off, the number of degrees of freedom is finite and generally proportional to the volume of the region considered. As soon as the effects of gravity are taken into account, such a picture is challenged by a number of semi-classical interrelated arguments: 

1) According to the \emph{holographic principle} \cite{hol}, the maximal 
entropy within a region is finite and proportional to the area -- rather than the volume -- of the region.
Although entanglement entropy very generally exhibits area scaling
\cite{scaling}, a localized maximally entropic state can be 
constructed  with entropy proportional to the volume; 
the holographic bound then forces 
to restrict \cite{yurt} the effective Fock space inside 
the region (see, however, \cite{hsu} for a different 
view).

2) Locality is questioned, at various levels, in several (e.g. \cite{solu}) proposed solutions of the \emph{black hole information-loss paradox}. There are also solid arguments \cite{gidd} against local QFT whenever describing field configurations that have non-negligeable effects on the background metric (e.g. when they tend to form closed trapped surfaces). 

3) It has been argued that the Hilbert space describing \emph{quantum gravity in asymptotically de Sitter} (dS) space is of finite dimension \cite{banks}.  Since the volume of the dS spatial sections changes with time, how can the fundamental degrees of freedom be local and separated by a fixed, short distance cut-off? 

4) More generally, there is an evident friction between
the naive picture of a fixed short distance cut-off and the dynamical nature of spacetime. In explicit contrast to the holographic principle, one may simply resolve to associate, to each finite volume of space, an infinite number of degrees of freedom and, potentially, an infinite entropy and an infinite amount of information.
But this seems hardly 
compatible \cite{hol} with the \emph{finiteness of the black hole entropy}, which is a well established result (see e.g. \cite{dam} for a review), explicitely worked out within string theory for certain extremal cases \cite{sen}. ``Where'' and how ``distributed'' are, then, 
the fundamental degrees of freedom? 

In the hope of possibly gaining some insight into these issues,
in this note I give a concise account of a preliminary attempt \cite{fedo}
to consider locality and spacetime itself under a novel perspective. 
As a matter of fact, any spacetime measurement is made by the mutual relations between objects, fields, particles etc\dots\  Any operationally meaningful assertion about spacetime is therefore intrinsic to the degrees of freedom of the matter (i.e. non-gravitational) fields and concepts such as ``locality'' and ``proximity'' should, at least in principle, be operationally definible entirely within the dynamics of the matter fields. In this respect, the usual approach of QFT follows quite an opposite route: the fundamental degrees of freedom are labelled already from the beginning by the spacetime points and locality is given \emph{a priori} as an attribute of the class of sub-systems that are to be considered. 

In quantum mechanics the Hilbert space of a composite system is the direct product of the Hilbert spaces of the components. Therefore, each possible way we can intend 
a given system (say ``the Universe'') as made of subsystems \cite{paolo1}, 
mathematically corresponds to a \emph{tensor product structure} (TPS) 
of its Hilbert space:
\begin{equation} \label{1}
\cH_{\rm Universe} \ = \ \cH_A \otimes \cH_B \otimes \cH_C  \otimes \dots \ .
\end{equation}
A decomposition into subsystems such as (\ref{1}) can be assigned without any reference to the locality or to the geometric properties of the components. Of course, 
an arbitrarily assigned TPS on a Hilbert space does not have much significance on its own: without any observable/operator of some definite meaning it 
looks impossible to extract any physical information about the system. However,
by knowing the dynamics inside $\cH_{\rm Universe}$, i.e. the unitary
operator $U(t_2,t_1)$ that, in the Schroedinger representation, evolves the state 
vectors according to $|\Psi(t_2)\ket = U(t_2,t_1)|\Psi(t_1)\ket$, we can, at least, follow 
the evolution of the \emph{correlations} between the subsystems. Correlations play a central role in Everett's view of quantum mechanics. In his seminal 
dissertation \cite{everett}, the relation between quantum correlations and mutual information is deeply exploited and measurements are consistently described as appropriate unitary evolutions that increase the degree of correlation between two subsystems: the ``measured'' and the ``measuring''. By taking Everett's view one can try to re-interpret the evolution of the system $|\Psi(t)\ket$ 
as measurements actually going on between the different parties $A$, $B$, $C$ etc\dots\ It is very 
compelling that locality itself and the usual local observables of direct physical interpretation may be eventually picked out within such an abstract scheme.

The issue that we try to address is to characterize
the class of TPSs that single out ``localized subsystems''; the ones, in other words, 
naturally associated with space-like separated regions of spacetime\footnote{Here $t$ plays 
the role of a global time parameter, and therefore the class of 
``local subsystems'' that we aim to define belongs to some given spacelike slicing. 
One should be able to recover \emph{a posteriori} the complete general covariance, as 
in the Hamiltonian formulation of field theories where a spatial slicing is initially required. 
The difference between the 
``external'' and ``inaccessible'' time parameter $t$ and the time as perceived by the observers 
is, on the other hand, discussed elsewhere \cite{fedo} (but see also \cite{time}).}.
As a source of correlation/information we only consider \emph{quantum entanglement}, which, 
for a bipartite system $AB$ in a pure state $|\Psi\ket$, is measured
by the \emph{von Neumann entropy} $S(A) = - {\rm Tr}_A (\rho_A \log_2 \rho_A)$, 
where $\rho_A = {\rm Tr}_B |\Psi\ket\bra\Psi|$.
The space of the TPSs is spanned by the elements of a group. 
By taking, for instance, $\cH_{\rm Universe}$ of dimensions $d^N$ and each of the $N$ 
subsystems (\ref{1}) of dimensions $d$, the group is $U(d^N)/U(d)^N$. 
The smaller the dimensions of the subsystems, the finer-grained the description that is given.

The most elementary type of spacetime relation that one can try to define intrinsically from
the dynamics of general subsystems is that of \emph{mutual spacetime coincidence}, what may be intuitively
viewed as ``being in the same place at the same time" or just ``having been in touch''. 
Inspired by the known local character of physical laws, we attempt to define 
coincidence by means of physical interactions, i.e. to define  two parties as 
``having been coincident" if they ``have physically interacted" with each other. 
By choosing the production of entanglement as a ``measure'' of interaction 
a sufficient condition for spacetime coincidence can be given:
\begin{quote}
\emph{Spacetime coincidence (sufficient condition)}: If before the instant $t_1$ the subsystems $A$ 
and $B$ are in a pure state \emph{i.e.} $S(A;t<t_1) = S(B;t<t_1) =0$ and, at a later time 
$t_2$ they are entangled, $S(A;t_2) = S(B;t_2) > 0$, without, during all the process, having mixed with anything else,
$S(AB;t<t_2) = 0$, then $A$ and $B$ \emph{have been coincident} with each other between $t_1$ and $t_2$.
\end{quote}
As a sufficient condition, the one above stated is rather strict: 
there are a number of physical situations that one would legitimately consider as ``spacetime coincidence relations'' but do not fit into the above definition. Most notably, two systems may interact with each other
while being already entangled with something else; i.e. not initially in a pure state. 
In this case, however, it is hard to give a quantitative definition of contiguity because a reliable measure of entanglement for multipartite systems is still a matter of debate. 

Note that, if $A$ and $B$ have been coincident, coincidence generally applies also to many 
other ``larger'' systems containing $A$ and $B$ as subsystems. 
The opposite can also be true. Say that $A$ is itself a composite system \emph{i.e.} $\cH_A = \cH_{A1}\otimes\cH_{A2} \otimes \dots$; we may discover that the sub-subsystem $A2$ is in fact ``responsible'' at a deeper level for the coincidence between $A$ and $B$. 
Again, the smaller the dimension of the systems which coincidence is recognized to apply, 
the more finely grained the spacetime description we are able to give.

The notion of coincidence can be applied to subsystems belonging to 
arbitrary TPSs and therefore possibly maximally unlocalized, such as, in ordinary QFT in flat space, those associated with the modes of given momentum. In the TPS of the localized subsystems, however, during an infinitesimal lapse of time, each subsystem creates new correlations with the smallest possible number of other subsystems: its ``neighbors''.  We argue therefore the following generic property of the TPSs that single
out localized subsystems: 
\begin{quote}
\emph{Locality Conjecture:} 
``Localized subsystems'' have the minimum tendency to create \emph{coincidence} relations with each other: the tensor product structure that singles out \emph{localized systems} is the one in which the entanglement of initially completely factorized states \emph{minimally} grows during time evolution. 
\end{quote}
In \cite{fedo} generic interacting second quantized models with a finite number of 
fermionic degrees of freedom have been considered. The symmetries of the Hamiltonian (in this case the conservation of number of particles) dramatically restrict the possible TPS choices. By applying the above conjecture to a one-dimensional 
Heisenberg spin chain and to two particles states the tensor product structure usually associated 
with ``position'' is recovered. While referring to that paper for more details, in the following we finally summarize how, according to this general approach, the relation between quantum degrees of freedom and spacetime regions should be re-considered.

The Hilbert space of a QFT with a short distance cut-off and formulated within a finite-size Universe can be written as $\cH = \otimes_{i = 1}^N \cH_i$, where $\cH_i$ is the Hilbert space of each of the bosonic or fermionic modes associated with the $N$ points of the lattice space. 
In order to describe a (connected or disconnected) ``region'' of space of given volume
we pick up a subset $R$ of the lattice space with a given number of points. The region is thus associated 
with a physical system of Hilbert space $\cH_R = \otimes_{i \in R} \cH_i$. 
There are only a finite number of possible ``regions'' because, due to the presence  of the cut-off, there are a total finite number of ``points''. 
For generic states, the instantaneous tendency of the subsystem $R$ to create new correlations with the ``outside'' is roughly proportional to its boundary, since only the points at the boundary contribute in this process. A sphere, according to the principle of minimal tendency to entanglement, is the ideal ``localized subsystem'' , since, for a given dimensionality/volume, it is the shape with minimum area. Our locality conjecture, therefore, applies already in this framework by excluding regions of space which are disconnected or very spread in only one or two directions.

We are arguing, on the other hand, that restricting only to partitions of the type $\cH_R = \otimes_{i \in R} \cH_i$ just reflects our prejudicial -- and operationally meaningless -- idea of a spacetime pre-existing and independent of the state of the fields. The tendency to entanglement should be minimized, for a given state, not only among the finite number of possible partitions of the $N$ points, but to a much wider \cite{fedo} 
(infinite) number of partitions (tensor product structures) of the total Hilbert space. 

\vspace{1cm}

{\bf Acknowledgements:}
It is a pleasure to thank Paolo Zanardi and the people of the quantum computation group
of the I.S.I foundation, Turin, for very useful discussions. I also thank Chris Clarkson for 
discussions as well as comments on the manuscript. This work has been supported by a 
Marie Curie Fellowship under contract number MEIF-CT-2004-502356.




\bibliography{sample}

\IfFileExists{\jobname.bbl}{}
 {\typeout{}
  \typeout{******************************************}
  \typeout{** Please run "bibtex \jobname" to optain}
  \typeout{** the bibliography and then re-run LaTeX}
  \typeout{** twice to fix the references!}
  \typeout{******************************************}
  \typeout{}
 }



\end{document}